\documentclass[12pt,a4paper,final]{iopart}
\usepackage{iopams}
\usepackage{bm}

\expandafter\let\csname equation*\endcsname\relax
\expandafter\let\csname endequation*\endcsname\relax

\usepackage{amsmath}
\usepackage{amssymb}
\usepackage{amsthm}
\usepackage[utf8]{inputenc}
\usepackage{graphicx}         
\usepackage{bm}               
\usepackage{color}
\usepackage{xcolor}
\usepackage{soul}
\usepackage{cite}

\usepackage[breaklinks=true,colorlinks=true,linkcolor=blue,urlcolor=blue,citecolor=blue]{hyperref}

\newcommand{\defeq}{\mathrel{\mathop:}=}

\newcommand{\prob}[2]{\ensuremath{P(#1|#2)}}
\renewcommand{\eqref}[1]{Eq. (\ref{#1})}

\begin{document}

\title{The $q$-canonical ensemble as a consequence of Bayesian superstatistics}

\author{Sergio Davis}
\address{Research Center in the Intersection of Plasma Physics, Matter and Complexity (P$^2$mc), Comisi\'on Chilena de Energía Nuclear, Casilla 188-D, Santiago, Chile}
\address{Departamento de Física, Facultad de Ciencias Exactas, Universidad Andres Bello. Sazi\'e 2212, piso 7, 8370136, Santiago, Chile}
\ead{sergio.davis@cchen.cl}

\begin{abstract}
Superstatistics is a generalization of equilibrium statistical mechanics that describes systems in nonequilibrium steady states. Among the possible superstatistical
distributions, the $q$-canonical ensemble (also known as Tsallis' statistics, and in plasma physics as Kappa distributions) is probably the most widely used,
however the current explanations of its origin are not completely consistent.

In this work it is shown that, under a Bayesian interpretation of superstatistics, the origin of the $q$-canonical
ensemble can be explained as the superstatistical distribution with maximum Shannon-Jaynes entropy under noninformative constraints. The $q$-canonical distributions
are singled out by the mathematical structure of superstatistics itself, and thus no assumptions about the physics of the systems of interest, or regarding their
complexity or range of interactions, are needed. These results support the thesis that the success of the $q$-canonical ensemble is information-theoretical in nature,
and explainable in terms of the original maximum entropy principle by Jaynes.
\end{abstract}

\maketitle

\section{Introduction}

Understanding the origin of non-canonical probability distributions ocurring in complex systems, particularly the $q$-canonical distributions, is one of the
remaining fundamental problems of nonequilibrium statistical mechanics. The $q$-canonical family of distributions (sometimes called Tsallis distributions)
generalize the well-known canonical distribution of Boltzmann-Gibbs statistical mechanics,
\begin{equation}
\prob{\bm{\Gamma}}{\beta} = \frac{\exp(-\beta H(\bm \Gamma))}{Z(\beta)},
\label{eq:canonical}
\end{equation}
for a system with microstates $\bm \Gamma \in U$ where $Z(\beta)$ is the partition function,
\begin{equation}
Z(\beta) \defeq \int_U d\bm{\Gamma} \exp(-\beta H(\bm \Gamma)),
\end{equation}
replacing it with the distribution
\begin{equation}
\label{eq:qens}
\prob{\bm{\Gamma}}{\beta_0, q} = \frac{\exp(-\beta_0 H(\bm \Gamma); q)}{Z_q(\beta_0)},
\end{equation}
where $\exp(x; q)$ is the $q$-exponential function, defined by
\begin{equation}
\exp(x; q) \defeq \big[1+(1-q)x\big]_+^{\frac{1}{1-q}},
\end{equation}
with $[x]_+ = \max(x, 0)$. Because $\exp(x; q) \rightarrow \exp(x)$ in the limit $q \rightarrow 1$, the canonical ensemble in \eqref{eq:canonical} is a
particular case of \eqref{eq:qens}.

In plasma physics, the $q$-canonical generalizations of the Maxwell-Boltzmann distribution of velocities are known as
Kappa distributions~\cite{Pierrard2010,Livadiotis2017,Espinoza2018, Ourabah2020b}, and have been found in space plasmas as well as in dense plasmas generated
in fusion experiments~\cite{Knapp2013,Klir2015}. Kappa distributions are commonly stated by making the correspondence~\cite{Livadiotis2009}
\begin{equation}
\kappa \defeq \frac{1}{q-1}
\end{equation}
with $\kappa > 0$ so that $q \geq 1$ and $q \rightarrow 1$ corresponds to $\kappa \rightarrow \infty$.

The issue of explaining the origin of the $q$-canonical distributions is still an open question, and in particular whether a generalization of the Boltzmann-Gibbs entropy, such
as the one proposed by Tsallis~\cite{Tsallis1988,Tsallis2009} is required for this. New mechanisms that give rise to $q$-canonical models, such as superstatistics~\cite{Beck2003,Beck2004} among others~\cite{Kaniadakis2001, Wang2002c} are challenging the need for generalized entropies.

Superstatistics in particular is firmly rooted on the rules of probability theory while still being compatible with Jaynes' original maximum entropy principle~\cite{Davis2020c}
and is therefore an appealing choice of framework. Moreover, superstatistics is especially suited to the language of Bayesian probability, as has been shown in several different
works~\cite{Sattin2006,Alves2016,Sattin2018,Davis2018,Sanchez2018,Umpierrez2021}. Interestingly, superstatistics has been shown to be the proper framework to describe the
velocity distribution of particles in collisionless plasmas~\cite{Davis2019b}.

In this work we show that the $\chi^2$-family of superstatistics, from which the $q$-canonical (Tsallis) distributions are obtained, arise naturally from the superstatistical
framework as a noninformative prior distribution of inverse temperature, that we recover from a standard application of the principle of maximum entropy and requiring only
compatibility with the structure of superstatistics itself. We therefore provide a strong argument for the ubiquity of the $q$-canonical ensembles solely based on the standard
principles of statistical mechanics and Bayesian/information theoretical reasoning, possibly enabling future insights regarding the meaning of $q$.

\section{A Bayesian retelling of the canonical ensemble}

The distribution of energies corresponding to \eqref{eq:canonical} is given by
\begin{equation}
\label{eq:canon_energy}
\prob{E}{\beta} = \frac{\exp(-\beta E)\Omega(E)}{Z(\beta)},
\end{equation}
with
\begin{equation}
\Omega(E) \defeq \int_U d\bm{\Gamma} \delta(E-H(\bm \Gamma)).
\end{equation}
the density of states. Upon simple inspection we notice an asymmetry between \eqref{eq:canonical} and \eqref{eq:canon_energy}, in that the energy distribution is weighted by
this density of states. In order to put both distributions in the same footing we will introduce the idea of \emph{Bayesian statistical ensembles}. These are
probability distributions of the form
\begin{equation}
\prob{\zeta}{\mathcal{R}, I_0} = \prob{\zeta}{I_0}\frac{\prob{\mathcal{R}}{\zeta, I_0}}{\prob{\mathcal{R}}{I_0}},
\label{eq:bayes}
\end{equation}
where $\zeta$ is a generic set of variables and $\mathcal{R}$ is some evidence or piece of information that updates the \emph{prior} distribution $\prob{\zeta}{I_0}$
to a \emph{posterior} distribution $\prob{\zeta}{\mathcal{R}, I_0}$ that incorporates $\mathcal{R}$. We see then that in \eqref{eq:canonical} we have omitted the \emph{prior}
distribution of microstates $\prob{\bm \Gamma}{I_0}$, which was originally taken as a constant by virtue of the postulate of \emph{a priori} equiprobability of microstates
and thus cancelled out. Including this \emph{prior} we should instead write
\begin{equation}
\label{eq:canonical_prior}
\prob{\bm{\Gamma}}{\beta, I_0} = \prob{\bm{\Gamma}}{I_0}\frac{\exp(-\beta H(\bm \Gamma))}{Z_0(\beta)}.
\end{equation}
as a replacement for \eqref{eq:canonical} where now
\begin{equation}
Z_0(\beta) \defeq \int_U d\bm{\Gamma} \prob{\bm{\Gamma}}{I_0}\exp(-\beta H(\bm \Gamma))
\end{equation}
is the proper partition function. Following this argument, the density of states has to be replaced by the prior distribution of energy
\begin{equation}
\prob{E}{I_0} \defeq \int_U d\bm{\Gamma}\prob{\bm \Gamma}{I_0}\delta(E-H(\bm \Gamma))
\label{eq:E_prior}
\end{equation}
so that we can write
\begin{subequations}
\begin{align}
\label{eq:canon_prior}
\prob{\bm{\Gamma}}{\beta, I_0} & = \prob{\bm{\Gamma}}{I_0}\frac{\exp(-\beta H(\bm \Gamma))}{Z_0(\beta)}, \\
\label{eq:canon_E_prior}
\prob{E}{\beta, I_0} & = \prob{E}{I_0}\frac{\exp(-\beta E)}{Z_0(\beta)}.
\end{align}
\end{subequations}

\noindent
We can see that the definition of $\prob{E}{I_0}$ is consistent with the proper normalization of \eqref{eq:canon_E_prior}, because
\begin{equation}
\begin{split}
Z_0(\beta) & = \int dE \prob{E}{I_0}\exp(-\beta E) \\
           & = \int dE\int_U d\bm{\Gamma}\prob{\bm{\Gamma}}{I_0}\delta(E-H(\bm \Gamma))\exp(-\beta E) \\
           & = \int_U d\bm{\Gamma} \prob{\bm{\Gamma}}{I_0}\exp(-\beta H(\bm \Gamma)).
\end{split}
\end{equation}

Notice that now \eqref{eq:canon_prior} and \eqref{eq:canon_E_prior} can both be derived using the maximization of a single form of entropy, namely the parameterization-invariant
Shannon-Jaynes entropy~\cite{Caticha2007},
\begin{equation}
\label{eq:entropy}
\mathcal{S}(I \rightarrow I_0) = -\int d\zeta \prob{\zeta}{I}\ln \left[\frac{\prob{\zeta}{I}}{\prob{\zeta}{I_0}}\right]
\end{equation}
under the constraint of fixed expected energy, where the generic variables $\zeta$ can be replaced by either the microstate variables $\bm\Gamma$ or the energy $E$. We notice also
that both \eqref{eq:E_prior} and \eqref{eq:canon_E_prior} are particular cases of the rule
\begin{equation}
\begin{split}
\prob{A = a}{S} & = \big<\delta(A-a)\big>_S \\
                & = \int_U\hspace{-5pt} d\bm{\Gamma} \prob{\bm{\Gamma}}{S}\delta(a-A(\bm \Gamma))
\end{split}
\end{equation}
for an arbitrary observable $A(\bm \Gamma)$ and state of knowledge $S$.

\section{Generalized ensembles and superstatistics the Bayesian way}

\noindent
Now, instead of the constraint of the expected energy we will determine the effect of a constraint on the form of the energy distribution $\prob{E}{S}$, which we will
assume is known and given by the function $p(E)$. That is, we will impose
\begin{equation}
\label{eq:con}
P(E|S) = \big<\delta(H-E)\big>_S = p(E)\quad \forall\; E \geq 0.
\end{equation}

\noindent
By maximizing the Shannon-Jaynes entropy in \eqref{eq:entropy} from the prior $P(\bm \Gamma|I_0)$ under the constraint in \eqref{eq:con} we obtain
\begin{equation}
\label{eq:ensemble_gamma}
\begin{split}
P(\bm \Gamma|S) & = \frac{1}{\eta}P(\bm \Gamma|I_0)\exp\left[-\int_0^\infty dE \mu(E)\delta(H(\bm \Gamma)-E)\right] \\
                & = \frac{1}{\eta}P(\bm \Gamma|I_0)\exp(-\mu(H(\bm \Gamma))),
\end{split}
\end{equation}
where $\mu(E)$ is the Lagrange multiplier function conjugate to $p(E)$. \eqref{eq:ensemble_gamma} can be rewritten as
\begin{equation}
\label{eq:rho}
P(\bm \Gamma|S) = P(\bm \Gamma|I_0)\rho(H(\bm \Gamma); S),
\end{equation}
with $\rho(E; S)$ a non-negative function of the energy known as the \emph{ensemble function}, in a manner that is consistent with Bayes' theorem in \eqref{eq:bayes}
if we use $\zeta=\bm{\Gamma}$ and let
\begin{equation}
\rho(H(\bm \Gamma); S) = \frac{P(\mathcal{R}|\bm \Gamma, I_0)}{P(\mathcal{R}|I_0)}.
\end{equation}

\noindent
The distribution of energy $P(E|S)$ that generalizes \eqref{eq:canon_energy} is then
\begin{equation}
\label{eq:Eprior}
\begin{split}
P(E|S) & = \Big<\delta(H-E)\Big>_S \\
       & = \int_U d\bm{\Gamma} P(\bm \Gamma|I_0)\rho(H(\bm \Gamma); S)\delta(H(\bm \Gamma)-E) \\
       & = P(E|I_0)\rho(E; S),
\end{split}
\end{equation}
again in agreement with Bayes' theorem in \eqref{eq:bayes} if $\zeta=E$ and we let
\begin{equation}
\rho(E; S) = \frac{P(\mathcal{R}|E, I_0)}{P(\mathcal{R}|I_0)}.
\end{equation}

While, in principle, every energy distribution $p(E)$ produces a valid steady-state ensemble in this way, we are interested in those compatible with the idea
of temperature in the framework of superstatistics. This is a particular class of models defined by a distribution of inverse temperature $P(\beta|S)$.
The central equation of superstatistics, as presented in recent works~\cite{Dixit2015,Davis2020}, is the joint distribution of microstates $\bm \Gamma$ and inverse temperature $\beta$,
which is given by the product rule of probability,
\begin{equation}
P(\bm \Gamma, \beta|S) = P(\bm \Gamma|\beta)P(\beta|S).
\end{equation}

\noindent
After using the marginalization rule we can extract the distribution of microstates,
\begin{equation}
P(\bm \Gamma|S) = \int_0^\infty\hspace{-5pt}d\beta P(\bm \Gamma, \beta|S) = \int_0^\infty\hspace{-5pt}d\beta P(\bm \Gamma|\beta)P(\beta|S),
\end{equation}
which upon replacing the canonical ensemble in \eqref{eq:canonical_prior} reads
\begin{equation}
P(\bm \Gamma|S) = P(\bm \Gamma|I_0)\int_0^\infty\hspace{-5pt}d\beta \left[\frac{\exp(-\beta H(\bm \Gamma))}{Z_0(\beta)}\right]P(\beta|S).
\end{equation}

\noindent
Direct comparison with \eqref{eq:rho} gives the ensemble function $\rho(E; S)$ as
\begin{equation}
\label{eq_rho_super}
\rho(E; S) = \int_0^\infty d\beta \left[\frac{P(\beta|S)}{Z_0(\beta)}\right]\exp(-\beta E),
\end{equation}
which tells us that $\rho(E; S)$ is the Laplace transform of the \emph{superstatistical weight function}
\begin{equation}
f(\beta; S) \defeq \frac{P(\beta|S)}{Z_0(\beta)}.
\end{equation}

From this point forward we will only consider superstatistical ensembles, and therefore we will omit the label $S$ on the functions $\rho(E)$ and $f(\beta)$ for simplicity.

The joint distribution of energy and inverse temperature $P(E, \beta|S)$ can be factorized using the product rule,
\begin{align}
P(E, \beta|S) = P(E|\beta)P(\beta|S),
\end{align}
and by replacing \eqref{eq:canon_energy} it can be written compactly as
\begin{equation}
P(E, \beta|S) = \exp(-\beta E)P(E|I_0)f(\beta).
\label{eq_joint_1}
\end{equation}

Note that this structure of a superstatistical joint distribution of $E$ and $\beta$ implies that the quantity \[P(E, \beta|S)\exp(\beta E)\] must always be separable as a
product of a function of energy and a function of (inverse) temperature.

\section{The fundamental inverse temperature}

For steady-state ensembles of the form in \eqref{eq:rho} we define the fundamental inverse temperature as the function
\begin{equation}
\beta_F(E; S) \defeq -\frac{\partial}{\partial E}\ln \rho(E; S).
\label{eq_beta_fund}
\end{equation}

\noindent
This function contains all the information about the ensemble, because integration of equation \eqref{eq_beta_fund} readily gives
\begin{equation}
\rho(E; S) = \rho(0; S)\exp\left(-\int_0^{E} dE'\beta_F(E'; S)\right),
\label{eq_rho_int}
\end{equation}
and thus, more importantly for us, recovering the correct form of $\beta_F$ is equivalent to recovering the correct ensemble function $\rho$. The fundamental inverse
temperature function is also usually easier to read and manipulate than the ensemble function. As the simplest example, the canonical ensemble at $\beta = \beta_0$ is
represented with a constant function $\beta_F(E; \beta_0) = \beta_0$.

In the case of superstatistics, there is a clear connection between the moments of the conditional distribution $P(\beta|E,S)$ and $\beta_F$. As shown in Ref.~\cite{Davis2022},
the $n$-th moment of this distribution is given by
\begin{equation}
\label{eq_beta_moments}
\big<\beta^n\big>_{E,S} = \frac{(-1)^n}{\rho(E)}\frac{\partial^n \rho(E)}{\partial E^n},
\end{equation}
which for $n = 1$ this yields $$\big<\beta\big>_{E,S} = -\frac{\partial}{\partial E}\ln \rho(E),$$
that is,
\begin{equation}
\big<\beta\big>_{E,S} = \beta_F(E; S).
\label{eq_beta_constraint}
\end{equation}

This equality gives meaning to the fundamental inverse temperature in superstatistics~\cite{Davis2019}, as the conditional mean inverse temperature for a fixed energy $E$. Using
$n = 2$ we obtain
\begin{equation}
\label{eq_beta_2}
\big<\beta^2\big>_{E,S} = \beta_F(E; S)^2 - {\beta_F}'(E; S)
\end{equation}
from which it follows that
\begin{equation}
\label{eq_betaf_fluc}
\big<(\delta \beta)^2\big>_{E,S} = \big<\beta^2\big>_{E,S} - \big<\beta\big>_{E,S}^2 = -{\beta_F}'(E),
\end{equation}
and therefore, for every valid superstatistics it must hold that
\begin{equation}
\label{eq:ineq_super}
{\beta_F}'(E) \leq 0.
\end{equation}

\section{The $q$-canonical ensemble}

\noindent
We will define the $q$-canonical ensemble by its fundamental inverse temperature function
\begin{equation}
\beta_F(E; \beta_0, q) = \frac{\beta_0}{1-(1-q)\beta_0 E},
\label{eq_betaf_super}
\end{equation}
where $q$ is traditionally known as the \emph{entropic index}, and $\beta_0$ an inverse temperature that serves as a scale parameter. Here it is immediately clear that
$q = 1$ recovers the canonical ensemble with constant fundamental inverse temperature
\begin{equation}
\beta_F(E; \beta_0, q=1) = \beta_0.
\end{equation}

\noindent
Integration of \eqref{eq_betaf_super} according to \eqref{eq_rho_int} gives the usual~\cite{Tsallis2009} ensemble function
\begin{equation}
\rho(E; \beta_0, q) = \rho(0)\big[1-(1-q)\beta_0 E\big]^{\frac{1}{1-q}},
\end{equation}
and because we can write
\begin{equation}
{\beta_F}'(E; \beta_0, q) = \frac{\beta_0^2(1-q)}{\big(1-(1-q)\beta_0 E\big)^2} = (1-q)\beta_F(E; \beta_0, q)^2
\end{equation}
we see that the inequality in \eqref{eq:ineq_super} implies $q \geq 1$.

As noted by Beck and Cohen~\cite{Beck2003}, this ensemble can be obtained in superstatistics from \eqref{eq_rho_super} with a weight function $f(\beta)$
of the form
\begin{equation}
f(\beta) = \frac{1}{\eta(\beta_0, q)}\exp\Big(-\frac{\beta}{\beta_0(q-1)}\Big)\:\beta^{\frac{1}{q-1}-1},
\label{eq_laplace_inv}
\end{equation}
referred to as the $\chi^2$-distribution. We can see this by considering the integral
\begin{align}
\int_0^\infty d\beta \exp\left(-\frac{\beta}{\beta_0(q-1)}\right)\exp(-\beta E)\beta^{\frac{1}{q-1}-1}
= \Gamma\left(\frac{1}{q-1}\right)\left[\frac{1+\beta_0 E (q-1)}{\beta_0(q-1)}\right]^{\frac{1}{1-q}}
\end{align}
and \eqref{eq:qens}, from which it follows that
\begin{equation}
\eta(q, \beta_0) = \Gamma\left(\frac{1}{q-1}\right)\Big[\beta_0(q-1)\Big]^{\frac{1}{q-1}}Z_q(\beta_0).
\end{equation}

The weight function $f(\beta)$ is proportional to a Gamma distribution with shape parameter $k = 1/(q-1)$ and scale parameter
$\theta = \beta_0(q-1)$. This distribution has a mean equal to $k\theta = \beta_0$ and a variance $k\theta^2 = \beta_0^2(q-1)$, and in the
limit $q \rightarrow 1$ we can see that the variance vanishes, thus $f(\beta)$ becomes proportional to $\delta(\beta-\beta_0)$ recovering the canonical ensemble.

\section{A first-principles derivation of the $q$-canonical ensemble}

Now we have all the elements needed to recover the $q$-canonical ensemble by its fundamental inverse temperature, namely \eqref{eq_betaf_super}, by maximizing the
Shannon-Jaynes entropy under constraints that only impose consistency with the superstatistical framework. More specifically, we will search for the maximum entropy joint distribution
$P(E, \beta|S)$ under the constraint that \eqref{eq_beta_constraint} holds for the whole range of allowed energies,
\begin{equation}
\big<\beta\big>_{E,S} = \beta_F(E; S) \qquad \forall\; E \geq 0.
\label{eq_C1}
\end{equation}

Additionally, because of the structure noted in \eqref{eq_joint_1}, we will require that the joint prior distribution of $E$ and $\beta$ is separable, i.e.,
\begin{equation}
P(E, \beta|I_0) = P(E|I_0)P(\beta|I_0).
\label{eq_C2}
\end{equation}

These are ``bare-bones'' constraints with almost no content. In fact, we have not included any information about a particular system, only requesting that the resulting model
must be compatible with superstatistics with a fundamental inverse temperature. However, in the following we will show that the least-biased pair of functions $\beta_F(E)$
and $P(\beta|I_0)$ that are consistent with these constraints and the structure of superstatistics are simply
\begin{equation}
\beta_F(E; \beta_0, q) = \frac{\beta_0}{1-(1-q)\beta_0 E},
\label{eq:betaf_result}
\end{equation}
that is, the $q$-canonical ensemble, and the inverse temperature prior
\begin{equation}
P(\beta|I_0) \propto \beta^{\frac{1}{q-1}-1}.
\label{eq:beta_result}
\end{equation}

\noindent
The prior distribution \eqref{eq:beta_result}, leading to \eqref{eq:betaf_result}, constitutes the main result of this work, and we proceed to prove it now. Recalling the
expectation identity (Eq. 14 of Ref.~\cite{Davis2020c}),
\begin{equation}
\Big<A\delta(B-b)\Big>_I = P(B = b|I)\big<A\big>_{B=b,I},
\end{equation}
valid for any pair of quantities $A$ and $B$, we have
\begin{equation}
\begin{split}
\Big<\beta\delta(H-E)\Big>_S & = \big<\beta\big>_{E,S}\;P(E|S) \\
& = \beta_F(E; S)\big<\delta(H-E)\big>_S.
\end{split}
\end{equation}

\noindent
Therefore, the constraint in \eqref{eq_C1} can be expressed as
\begin{equation}
\Big<\Big[\beta-\beta_F(E; S)\Big]\delta(H-E)\Big>_S = 0 \qquad\forall\; E \geq 0.
\label{eq_constraint_new}
\end{equation}

From this point on we will simplify the notation to write $\beta_F(E)$, leaving implicit the label $S$ of the steady state parameters. Maximizing the Shannon-Jaynes
entropy in the $(E, \beta)$ space with the constraint in \eqref{eq_constraint_new} and the prior $P(E, \beta|I_0)$ we obtain
\begin{align}
\label{eq_joint_2}
P(E, \beta|S) & = \frac{1}{\zeta[\lambda]} P(E, \beta|I_0)\exp\Big(-\int_0^\infty dE' \lambda(E')\big[\beta-\beta_F(E')\big]\delta(E-E')\Big) \nonumber \\
              & = \frac{P(E|I_0)P(\beta|I_0)}{\zeta[\lambda]}\exp\Big(-\lambda(E)\big[\beta-\beta_F(E)\big]\Big),
\end{align}
where $\lambda(E)$ is a Lagrange multiplier function to be determined, precisely the conjugate to the fundamental inverse
temperature $\beta_F(E)$ that fixes the constraint in \eqref{eq_C1}.

Note that, at this point, the functions $\lambda(E)$, $\beta_F(E)$ and $P(\beta|I_0)$ are all unknown, and in order to determine them, we impose consistency between
\eqref{eq_joint_1} and \eqref{eq_joint_2}. On the one hand, from \eqref{eq_joint_1} we obtain that
\begin{equation}
\frac{\partial}{\partial E}\ln P(E, \beta|S) = -\beta + \frac{\partial}{\partial E}\ln P(E|I_0).
\end{equation}
while \eqref{eq_joint_2} produces
\begin{equation}
\begin{split}
\frac{\partial}{\partial E}\ln P(E, \beta|S) = -\beta\lambda'(E) + \frac{\partial}{\partial E}\Big(\lambda(E)\beta_F(E)\Big) \\
 + \frac{\partial}{\partial E}\ln P(E|I_0).
\end{split}
\end{equation}

\noindent
As both must be true for all $\beta$ and $E$, we have that
\begin{equation}
-\beta = -\beta\lambda'(E) + \frac{\partial}{\partial E}\Big(\lambda(E)\beta_F(E)\Big),
\label{eq:master}
\end{equation}
and by comparing powers of $\beta$ on both sides it follows that $\lambda'(E) = 1$, therefore
\begin{equation}
\lambda(E) = E + \lambda_0
\end{equation}
where $\lambda_0$ is an integration constant, and
\begin{equation}
\frac{\partial}{\partial E}\Big(\lambda(E)\beta_F(E)\Big) = 0
\end{equation}
from which it follows that
\begin{equation}
\beta_F(E) = \frac{C}{E+\lambda_0}
\label{eq:betaf_uno}
\end{equation}
where $C$ is an additional integration constant. By introducing new parameters
\begin{equation}
\beta_0 \defeq \frac{C}{\lambda_0}, \quad q \defeq 1+\frac{1}{C}
\end{equation}
we recognize $\beta_F(E)$ in \eqref{eq:betaf_uno} as the fundamental inverse temperature of the $q$-canonical ensemble,
\begin{equation}
\beta_F(E) = \frac{\beta_0}{1-(1-q)\beta_0 E}.
\end{equation}

Now all that remains is to show that there is a prior $P(\beta|I_0)$ compatible with $\lambda(E)$ and $\beta_F(E)$. Replacing $\lambda(E)$ and $\beta_F(E)$ in
equation \eqref{eq_joint_2} we obtain
\begin{equation}
P(E, \beta|S) = \left[\frac{e^{1/(q-1)}}{\zeta(q, \beta_0)}P(\beta|I_0)\exp\Big(-\frac{\beta}{\beta_0(q-1)}\Big)\right]\exp(-\beta E)P(E|I_0),
\end{equation}
but agreement with \eqref{eq_joint_1} requires the quantity in square brackets to be equal to $f(\beta)$, therefore we have
\begin{equation}
f(\beta) = \frac{e^{1/(q-1)}}{\zeta(q, \beta_0)}P(\beta|I_0)\exp\Big(-\frac{\beta}{\beta_0(q-1)}\Big).
\label{eq:beta_prior}
\end{equation}

\noindent
Finally, because our fundamental inverse temperature $\beta_F(E)$ corresponds to the $q$-canonical ensemble, the function $f(\beta)$ in \eqref{eq:beta_prior} must agree
(up to a constant factor) with $f(\beta)$ in \eqref{eq_laplace_inv} for the $\chi^2$-distribution. This forces our prior for $\beta$ to be
\begin{equation}
P(\beta|I_0) \propto \beta^{\frac{1}{q-1}-1},
\label{eq:beta_improper}
\end{equation}
and at this point we can state our result as the fact that the constraints in \eqref{eq_C1} and \eqref{eq_C2} lead to
\begin{equation*}
f(\beta) = \frac{1}{\eta(\beta_0, q)}\exp\Big(-\frac{\beta}{\beta_0(q-1)}\Big)\:\beta^{\frac{1}{q-1}-1}
\end{equation*}
as the least-biased model, which is $\chi^2$-superstatistics. We can also interpret the result in \eqref{eq:beta_improper} as a noninformative prior~\cite{Berger2013} for
$\beta$ that does not refer to any particular system. Because $q \geq 1$, this prior is improper, i.e. not normalizable by itself, but
this is a common situation for noninformative priors in Bayesian inference, as is the case for instance for the Jeffreys' prior~\cite{Jaynes2003,Sivia2006}. The entropic index
$q$ can then be understood as the shape parameter of this noninformative prior.

\section{Concluding remarks}

A Bayesian view of statistical mechanics that includes superstatistics at its core, and uses the concept of fundamental inverse temperature, is by itself capable of
producing the main ensemble of Tsallis' nonextensive statistics. The entropic index $q$ appears to be the shape parameter of the noninformative prior distribution of
inverse temperature, and its origin does not invoke the use of any entropy other than the Boltzmann-Gibbs entropy, or rather, the parameterization-invariant Shannon-Jaynes
entropy. These results suggest that the $q$-canonical ensemble is not the posterior distribution of a generalized inference procedure using constraints but, rather, a
(noninformative) prior distribution which is built into superstatistics and should be its default. This is consistent with the derivation in Ref.~\cite{Davis2019} that
produces the $q$-canonical ensemble from invariance requirements on the fundamental inverse temperature $\beta_F(E)$, and in both cases the parameters $q$ and $\beta_0$
are not fixed \emph{a priori}.

 In view of the recent finding that particles in collisionless plasmas should follow superstatistical velocity distributions~\cite{Davis2019b}, the results of this work
provide additional arguments for the presence of the particular kind of superstatistics ($\chi^2$) that recovers the Kappa distribution.

\section*{Acknowledgements}

SD thankfully acknowledges financial support from ANID FONDECYT 1220651.

\section*{References}


\end{document}